\documentclass[11pt,a4paper]{article}
\pdfoutput=1

\usepackage{bm}
\usepackage{amsmath,amssymb}
\usepackage{cite}

\usepackage{graphicx}
\usepackage{color}
\usepackage{upgreek}

\usepackage{hyperref} %%put this package declaration last to make work since it redefines things!!%%
\hypersetup{colorlinks=true, citecolor=blue, filecolor=black, linkcolor=blue, urlcolor=blue, pdfpagemode=UseNone}
\numberwithin{figure}{section}
\numberwithin{equation}{section}

\newcommand{\be}{\begin{equation}}
\newcommand{\ee}{\end{equation}}
\newcommand{\bea}{\begin{eqnarray}}
\newcommand{\eea}{\end{eqnarray}}
\def\beal#1\eeal{\begin{align}#1\end{align}}   %one & only for aligning
\def\besp#1\eesp{\begin{multline}#1\end{multline}} %split an equation with first line left aligned & later right aligned

\usepackage[normalem]{ulem}

\newcommand{\cL}{\mathcal{L}}

\newcommand{\ph}{\varphi}
\newcommand{\vp}{\varphi}

\newcommand{\dd}[2]{\delta_{\!\phantom{(} #1}^{\!(#2)}\!(\ph)}

\newcommand{\h}{h }

\newcommand\ie{\textit{i.e.}\ }
\newcommand\eg{\textit{e.g.}\ }

\newcommand{\half}{\tfrac{1}{2}}

\begin{document}

\begin{titlepage}
%\begin{flushright}
%%{\tt hep-ph/yymmnn}
%{\tt SHEP xx-xx}
%\end{flushright}

\begin{center}
{\huge \bf Perturbatively renormalizable quantum gravity}

%Renormalizable quantum gravity and diffeomorphism invariance}

%of the conformal sector in quantum gravity}
%Relevant directions for the conformal factor in perturbative quantum gravity 
%\emph{or: Through the conformal factor to(wards) perturbatively renormalizable quantum gravity }} 
%\vskip.3cm
%{\huge \bf  and etc} 
\end{center}
\vskip1cm

%\title{xxx}
%\author{Tim R. Morris}

\begin{center}
{\bf Tim R. Morris}
\end{center}

%\affiliation{
\begin{center}
{\it STAG Research Centre \& Department of Physics and Astronomy,\\  University of Southampton,
Highfield, Southampton, SO17 1BJ, U.K.}\\
\vspace*{0.3cm}
{\tt  T.R.Morris@soton.ac.uk}
\end{center}

\abstract{The Wilsonian renormalization group (RG) requires Euclidean signature. The conformal factor of the metric then has a wrong-sign kinetic term, which has a profound effect on its RG properties. In particular around the Gaussian fixed point, it supports a Hilbert space of renormalizable interactions involving arbitrarily high powers of the gravitational fluctuations. These interactions are characterised by being exponentially suppressed for large field amplitude, perturbative in Newton's constant but non-perturbative in Planck's constant. By taking a limit to the boundary of the Hilbert space, diffeomorphism invariance is recovered whilst retaining renormalizability. Thus the so-called conformal factor instability points the way to constructing a perturbatively renormalizable theory of quantum gravity.}

% the limits are sufficiently flexible 
%
%we show that it is possible to take a limit to the boundary of $\Ll$ that is sufficiently flexible so as to retain renormalizability and allow diffeomorphism invariance to be recovered.}

\vskip4cm
\begin{center}
\textit{Essay written for the Gravity Research Foundation 2018 Awards for Essays on Gravitation.}\\
\textit{Submission date: March 30, 2018} %\today}
\end{center}

\end{titlepage}

If one follows the by--now--standard procedures of perturbative quantum field theory, such as those that make up the highly successful Standard Model of particle physics, 
then one finds that quantum gravity suffers from the problem that it is not perturbatively renormalizable \cite{tHooft:1974toh}.

This single stumbling block has spawned many fantastic ideas for combining quantum mechanics and gravity, none of which however has lead to a completely convincing theory. Indeed the very fact that so many attempted routes continue to be pursued, is testament to the truth that none has proved wholly persuasive. 

In gravity the natural coupling constant is $\kappa=2/M$, where $\kappa^2 = 32\pi G/\hbar c$, and $M$ is the reduced Planck mass. Given that $\kappa$ has negative mass dimension, perturbative non-renormalizability is expected from simple power counting arguments.  We claim however that perturbatively renormalizable quantum gravity may exist, and that properties already inherent to the Einstein-Hilbert action tell us how it should be constructed. Although this theory is perturbative in $\kappa$, it is non-perturbative in Planck's constant, $\hbar$. 

To understand why there is this possibility, one needs to work with the deeper understanding of renormalization afforded by the Wilsonian RG \cite{Wilson:1973}. An essential ingredient for this concept, is the Kadanoff blocking \cite{Kadanoff:1966wm}, where one integrates out degrees of freedom at short distances to obtain effective short range interactions. Thus one must work in Euclidean signature, so that short distance really does imply short range. This leads to the infamous ``conformal factor instability''. The Einstein-Hilbert action, $S_{EH}$, is the integral over the scalar curvature. Since there exist Euclidean manifolds with arbitrarily large curvature of either sign, the action cannot be bounded below.\footnote{In fact it is large positive curvature that is the problem.} It means that the functional integral,
$$
\mathcal{Z} = \int\!\! \mathcal{D}g_{\mu\nu}\ {\rm e}^{-S_{EH}/\hbar}\,,
$$
from which we would hope to construct the theory, is more than usually ill-defined. 

To construct the theory as a genuine quantum field theory, we must be able to take a continuum limit. Let $\Lambda$ be the mass-energy scale associated to the effective interactions. This is the inverse of the distance scale, and also the scale of the effective cutoff built into the Wilsonian effective action. At its simplest, taking a continuum limit means firstly finding a fixed point action under the RG, and then perturbing this action by adding all \emph{and only} the relevant interactions. (Universality follows from the fact that the irrelevant interactions cannot keep their own couplings.) The continuum limit is reached by tuning the corresponding relevant couplings as $\Lambda\to\infty$, in a way that is determined by the RG itself.

For a RG fixed point to exist, the manifold must itself look the same at any scale. That tells us to work with fluctuations on flat $\mathbb{R}^4$. Expressing those fluctuations as 
$$
g_{\mu\nu} = \delta_{\mu\nu}+\kappa H_{\mu\nu}\,,\qquad\text{where}\qquad H_{\mu\nu} = h_{\mu\nu} +\half\,\vp \,\delta_{\mu\nu}\,,
$$ 
$h_{\mu\nu}$ being the traceless part, and $\vp$ the traceful (conformal factor) part, and choosing a Feynman -- De Donder gauge, the problem is clearly visible already for free gravitons and isolated in the kinetic term for $\ph$ which has the wrong sign for convergence of the partition function:
$$
\cL^{\rm kinetic}_{EH} = \frac12 \left(\partial_\lambda \h_{\mu\nu}\right)^2 -\frac12 \left(\partial_\lambda\ph \right)^2\,.
$$
Previously the problem has been dealt with by continuing the conformal factor functional integral along the imaginary axis: $\ph\mapsto i\ph$ \cite{Gibbons:1978ac}. Instead we keep the instability and find that it has a profound effect on the Wilsonian RG. 

In the Wilsonian viewpoint, the free field action is the Gaussian fixed point, and from this one reads off  that both $h_{\mu\nu}$ and $\vp$ are to be regarded as having unit dimension.\footnote{RG scaling dimension, which is equal to the mass-dimension at the Gaussian fixed point.} The (marginally) relevant couplings are just those we then find have non-negative dimension. In the language of perturbation theory, they are the renormalizable couplings. Indeed the power counting arguments mentioned earlier also tell you that renormalizable couplings should have non-negative dimension.

With the right sign kinetic term, for example for $h_{\mu\nu}$ (also for $\vp$ if we continue it along the imaginary axis), perturbing around the Gaussian fixed point gives polynomial interactions. More precisely,  eigen-perturbations are Hermite polynomials, which form a complete orthonormal basis from which one can build any effective interaction in this Hilbert space providing it grows slow enough with large amplitude that it is square integrable under a Sturm-Liouville weight function:
$$
\omega_\Lambda(h) =\exp\left(-\frac{a^2 h^2_{\mu\nu}}{\hbar\Lambda^2}\right)\,,
$$
where $a$ is a non-universal constant \cite{Morris:1996nx}. The dimension of the coupling is set by the highest power in the polynomial (the lower powers generated by tadpole quantum corrections $\sim\hbar\Lambda^2$). The problem of perturbative renormalizability in this language is simply that the interactions you get from the Einstein-Hilbert action take the form\footnote{Indices suppressed; $n$ (later $m$) a non-negative integer; $\partial$ a space-time differential.} $h^n \partial h\partial h$, so that all the corresponding couplings have negative dimension. Thus the only continuum limit that can be formed is the one for free gravitons. 

For $\vp$ however, one finds that the wrong sign kinetic term also changes the sign in the weight function \cite{Dietz:2016gzg,Morris:2018mhd}:
$$
\omega_\Lambda(\vp) = \exp\left(+\frac{a^2 \vp^2}{\hbar\Lambda^2}\right)\,.
$$
For universality and the continuum limit,  we need
to be able to parametrise perturbations using couplings, \ie we need a Hilbert space structure. This tells us we cannot use polynomials now, since polynomials are not square integrable under $\omega_\Lambda(\vp)$. Instead the Hilbert space is spanned by the following orthonormal set of eigen-operators:
$$
\dd{\Lambda}{n} = \frac{\partial^n}{\partial\vp^n}\, \dd{\Lambda}{0}\,, \qquad{\rm where}\qquad \dd{\Lambda}0 = \frac{a}{\Lambda}\,\exp\left(-\frac{a^2\vp^2}{\hbar\Lambda^2}\right)\,.
$$
Since these ``$\delta$-operators'' have dimension $-1\!-n$, if they appeared on their own in the effective action, the corresponding couplings $g_n$ would have dimension $5\!+\!n$. They would thus form an infinite tower of interactions, whose couplings all have positive dimension, and are thus all renormalizable. Also in dramatic contrast to the polynomials, these operators are non-perturbatively quantum, \ie have no Taylor expansion in $\hbar$. 

Furthermore, purely $h$ interactions are now ruled out since a  constant in $\vp$ is also not integrable under  $\omega_\Lambda(\vp)$. Instead the most general interaction that is allowed can be expressed as $\sigma(h,\partial,\partial\vp) f_\Lambda(\vp)$, where $\sigma$ is a monomial containing powers of $h$, $\partial^m h$, and $\partial^m\vp$, and 
$$
f_\Lambda(\vp) = \sum^\infty_{n=0} g_n \dd{\Lambda}{n}\,.
$$
Here again the $g_n$ are the corresponding couplings, now subsumed in a \emph{coefficient function}. 
At first sight, factors of $\vp$ should be allowed in $\sigma$, and further factors of $\dd{\Lambda}{m}$ could appear, but together with $f_\Lambda$ these give functions of $\vp$ that are square integrable under $\omega_\Lambda(\vp)$ and thus by the Hilbert space property, can be re-expanded as a linear combination of the $\delta$-operators.

To build the quantum field theory, we need to include all the couplings with positive dimension, discarding those low-$n$ couplings which have negative dimension when $\sigma$ is taken into account. In this way we create renormalizable interactions involving arbitrarily high powers in $\sigma(h,\partial,\partial\vp)$.

But to be gravity, we must be able to choose couplings that allow diffeomorphism invariance. This means restoring BRST invariance on removing the cutoff, adapting standard methods, but also dealing with breaking by the coefficient functions \cite{Morris:2018}.
For the latter, consider the interaction of gravity with matter. At first order in $\kappa$ this must now take the form:
$$
\cL^{\rm int} \sim T^{\mu\nu} f_\Lambda(\vp) H_{\mu\nu}\,,
$$
where $T_{\mu\nu}$ is the stress-energy tensor, and $\kappa$ has been absorbed into the couplings $g_n$. Their dimension is therefore $[g_n]=n$, and thus $\cL^{\rm int}$ is renormalizable in all its couplings (except possibly the marginal $g_0$). This interaction should be invariant under linearised diffeomorphism invariance, $\delta H_{\mu\nu} = 2 \partial_{(\mu} \xi_{\nu)}$. Without the $f_\Lambda$, this is true by integration by parts, using energy-momentum conservation: $\partial_\mu T^{\mu\nu}=0$. But with $f_\Lambda$ in the way, diffeomorphism invariance is lost. 

However $f_\Lambda(\vp)$ can be any function that is square integrable under $\omega_\Lambda(\vp)$, and we will send $\Lambda\to\infty$ when we form the continuum limit. In this way we can choose the $g_n$ so that at any fixed value of $\vp$, $f_\Lambda\to\kappa$ in this limit. For example this is achieved by setting $f_\Lambda = \kappa/\omega_\Lambda(\vp)$. For higher dimension interactions where $m$ low-$n$ couplings must be set to zero to maintain renormalizability, we incorporate $m$ parameters \eg $f_\Lambda=\kappa\sum_{n=0}^m a_n/\omega^{\gamma_n}_\Lambda$, $\gamma_n>\half$ and $\sum_{n=0}^m a_n=1$.

Thus it seems we have all the elements necessary to construct quantum gravity as a genuine quantum field theory, and moreover the properties required were there all along in the structure of the action for General Relativity.  The resulting quantum field theory however is radically different from its Standard Model cousins.  Although it can be treated perturbatively in $\kappa$, it is non-perturbative in $\hbar$, which means that perturbative corrections are computed only by summing over all loops simultaneously,  and it is inherent to the construction that diffeomorphism invariance is recovered only after the passage to the continuum limit.\footnote{There are several more striking and important properties which we do not have room to convey \cite{Morris:2018mhd,Kellett:2018loq}.}

This should be contrasted with the so many other attempted routes to quantum gravity. Meanwhile experimental and theoretical progress has resulted in an increasing number of questions 
for such a theory: what is dark energy (or looks like it)? what is dark matter (or looks like it)? how do black holes emit Hawking radiation but stay consistent with unitarity? what drives or mimics inflation in the early universe? what ensured that the universe, inflated or otherwise, started sufficiently homogeneous? 

What physics will this theory unveil? We do not know yet, but we are about to find out.

%\bigskip\bigskip

\section*{Acknowledgments}
I acknowledge support from the Leverhulme Trust and Royal Society as a Royal Society Leverhulme Trust Senior Research Fellow, and from STFC Consolidated Grant ST/P000711/1.

%\newpage
%\mbox{}
%\newpage

%\bibliographystyle{hunsrt}
%\bibliography{references} %%from now on (14/7/15) this is the global references list!

\end{document}